\documentclass[aps,prd,twocolumn,superscriptaddress]{revtex4-2}

\usepackage{graphicx}
\usepackage{xspace}
\usepackage{xcolor}
\usepackage{hyperref}

\newcommand{\dcc}{LIGO-P1900183-v7}
\newcommand{\aligo}{Advanced LIGO (aLIGO)\gdef\aligo{aLIGO\xspace}\xspace}

\newcommand{\gw}[1][]{Gravitational wave#1 (GW#1)\renewcommand{\gw}[1][]{GW##1\xspace}\xspace}
\providecommand{\ns}[1][]{Neutron star#1 (NS#1)\renewcommand{\ns}[1][]{NS##1\xspace}\xspace}
\renewcommand{\ns}[1][]{Neutron star#1 (NS#1)\renewcommand{\ns}[1][]{NS##1\xspace}\xspace}
\newcommand{\cw}[1][]{continuous wave#1 (CW#1)\renewcommand{\cw}[1][]{CW##1\xspace}\xspace}
\newcommand{\tcw}[1][]{`transient Continuous Wave#1' (tCW#1)\renewcommand{\tcw}[1][]{tCW##1\xspace}\xspace}
\newcommand{\gpu}[1][]{graphics processing unit#1 (GPU#1)\renewcommand{\gpu}[1][]{GPU##1\xspace}\xspace}
\newcommand{\Fstat}[1][]{$\F$-statistic#1\xspace}
\newcommand{\snr}[1][]{signal-to-noise ratio#1 (SNR#1)\renewcommand{\snr}[1][]{SNR##1\xspace}\xspace}
\newcommand{\psd}{power spectral density (PSD)\renewcommand{\psd}{PSD\xspace}\xspace}
\newcommand{\sft}[1][]{Short Fourier Transform#1 (SFT#1)\renewcommand{\sft}[1][]{SFT##1\xspace}\xspace}
\newcommand{\ul}[1][]{upper limit#1 (UL#1)\renewcommand{\ul}[1][]{UL##1\xspace}\xspace}
\newcommand{\velapsr}{J0835--4510\xspace}
\newcommand{\crabpsr}{J0534+2200\xspace}

\newcommand{\F}{\mathcal{F}}
\newcommand{\Tsft}{T_\mathrm{SFT}}
\newcommand{\Nsft}{N_\mathrm{SFT}}
\newcommand{\Tobs}{T_\mathrm{obs}}

\newcommand{\fspin}{f_\mathrm{spin}}
\newcommand{\fgw}{f}
\newcommand{\fdot}{{\dot{\fgw}}}
\newcommand{\fddot}{\ddot{\fgw}}
\newcommand{\Amp}{\mathcal{A}}
\newcommand{\Dop}{\lambda}
\newcommand{\TP}{\mathcal{T}}
\newcommand{\Ntemp}{N_{\Dop}}
\newcommand{\Ntrans}{N_{\TP}}
\newcommand{\Neff}{N^\mathrm{eff}}

\newcommand{\Ntranseff}{N_{\TP}^\mathrm{eff}}
\newcommand{\Tgl}{T_\mathrm{glitch}}
\newcommand{\BSGL}{B_\mathrm{SGL}}
\newcommand{\Ftho}{\F_*^{(0)}}
\newcommand{\win}{\varpi}
\newcommand{\Fmn}{\F_{mn}}
\newcommand{\Tref}{T_\mathrm{ref}}
\newcommand{\days}{\mathrm{days}}
\newcommand{\Psft}{\mathcal{P}} 
\newcommand{\Sn}{S_{\mathrm{n}}}	
\newcommand{\SnXal}{\Sn^{X\alpha}}		
\newcommand{\Izz}{\mathcal{I}}
\newcommand{\pdet}{p_\mathrm{det}}
\newcommand{\pfa}{p_\mathrm{fa}}
\newcommand{\hul}{h_0^{90\%}}
\newcommand{\Egl}{E_\mathrm{glitch}}
\newcommand{\Isuper}{\Izz_\mathrm{s}}
\newcommand{\Inormal}{\Izz_\mathrm{c}}
\newcommand{\lag}{\Delta_{\mathrm{s-c}}}

\begin{document}

\preprint{\dcc}

\title{First search for long-duration transient gravitational waves after glitches in the Vela and Crab pulsars}


\newcommand{\glasgow}{\affiliation{University of Glasgow, School of Physics and Astronomy, Kelvin Building, Glasgow G12 8QQ, Scotland, United Kingdom}}
\newcommand{\michigan}{\affiliation{University of Michigan, Ann Arbor, MI 48109, USA}}
\newcommand{\jodrell}{\affiliation{Jodrell Bank Centre for Astrophysics, School of Physics and Astronomy, University of Manchester, Manchester M13 9PL, UK}}
\author{David Keitel}
\email[]{david.keitel@ligo.org}
\affiliation{University of Portsmouth, Institute of Cosmology and Gravitation, Portsmouth PO1 3FX, United Kingdom}
\glasgow
\author{Graham Woan}
\author{Matthew Pitkin}
\glasgow
\author{Courtney Schumacher}
\affiliation{Southern Methodist University, Dallas TX 75205, United States of America}
\author{Brynley Pearlstone}
\glasgow
\author{Keith Riles}
\michigan
\author{Andrew G. Lyne}
\jodrell
\author{Jim Palfreyman}
\affiliation{Department of Physical Sciences, University of Tasmania, Private Bag 37, Hobart, Tasmania 7001, Australia}
\author{Benjamin Stappers}
\author{Patrick Weltevrede}
\jodrell

\date{18 September 2019 [\dcc]}

\begin{abstract}
\gw[s] can offer a novel window into the structure and dynamics of neutron stars.
Here we present the first search for long-duration quasi-monochromatic \gw transients
triggered by pulsar glitches.
We focus on two glitches observed
in radio timing of the Vela pulsar (PSR~\velapsr) on 12 December 2016
and the Crab pulsar (PSR~\crabpsr) on 27 March 2017,
during the Advanced LIGO second observing run (O2).
We assume the GW frequency lies within a narrow band around twice the
spin frequency as known from radio observations.
Using the fully-coherent transient-enabled \Fstat method,
we search for
transients of up to four months in length.
We find no credible \gw candidates for either target,
and through simulated signal injections we set 90\% upper limits
on (constant) \gw strain as a function of transient duration.
For the larger Vela glitch,
we come close to beating an indirect upper limit
for when the total energy liberated in the glitch
would be emitted as \gw[s],
thus demonstrating that similar post-glitch searches
at improved detector sensitivity
can soon yield physical constraints on glitch models.
\end{abstract}

\maketitle

\section{\label{sec:intro}Introduction}

\ns[s] provide a rich astrophysical laboratory for nuclear physics at extreme densities.
\gw[s] can contribute to probing \ns structure and dynamics
through observations of binary mergers like GW170817~\cite{TheLIGOScientific:2017qsa},
but also by searching for signals
from individual, rapidly spinning objects~\cite{Glampedakis:2017nqy}.
One of the most prominent dynamic features of individual \ns[s]
is the presence of timing glitches (sudden spin-up events) in pulsars~\cite{Fuentes:2017bjx}.
Glitches are considered promising probes of the \ns interior~\cite{Link:1992mdl,Link:2000mu,Haskell:2017ngx}
and possible emitters of detectable \gw signals~\cite{vanEysden:2008pd,Bennett:2010tm,Prix:2011qv,Melatos:2015oca,Singh:2016ilt}.

Despite 50 years of glitch observations and a lot of productive model development~\cite{Haskell:2015jra},
the mechanism (or mechanisms) behind glitches are not well understood.
This is where detecting transient \gw signals during or following a glitch could yield valuable insights,
as they would directly measure changes in the quadrupole moment of the \ns.

So far, the only dedicated analysis of LIGO-Virgo~\cite{TheLIGOScientific:2014jea,TheVirgo:2014hva} data
that touched on this topic was
a search for short ($\sim\mathcal{O}$(s)) signals from a 2006 glitch of the Vela pulsar in initial LIGO data~\cite{Abadie:2010sf}.
Targeted searches for \cw signals from known pulsars
(most recently~\cite{Abbott:2017ylp,Abbott:2017cvf,Abbott:2019ztc,Abbott:2019bed})
have also taken observed glitches into account as breaks in the pulsar ephemerides,
but these always focus on persistent signals lasting for the full observation times before/after the glitch.
Intermediate-duration transient searches have also been performed on magnetar bursts
(most recently in~\cite{Abbott:2019dxx}
and to look for a post-merger remnant of GW170817~\cite{Abbott:2017dke,Abbott:2018hgk},
but  for those targets
the emission mechanisms and parameter space are very different
than for pulsar glitches.

Here we perform a dedicated transient analysis
of \gw data from the \aligo second observing run
(O2, December 2016 to August 2017),
searching for transient \gw signals lasting hours to months
after two glitches observed from
the Vela pulsar (PSR~\velapsr) on 12 December 2016
(MJD\,$57734.484991\pm0.000029$~\cite{Palfreyman:2016gli,Sarkissian:2017awh,Palfreyman:2018gli})
and the Crab pulsar (PSR~\crabpsr) on 27 March 2017
(MJD\,$57839.92\pm0.06$,~\cite{JodrellCrab:2018,EspinozaGlitchCatalogue}).
Due to their relatively young ages and close distances,
these were the first two pulsars for which
the indirect spin-down upper limit on \cw emission~(see e.g. Sec.~2.3 of~\cite{Prix:2009oha})
was beaten~\cite{Abbott:2008fx,Abadie:2011md}.
They are not the only
known pulsars to have glitched during O2,
and targeting a larger sample will be interesting in the future.
Sec. 2.2 of~\cite{Abbott:2019ztc} gives an overview of known pulsars
accessible to LIGO-Virgo searches,
though not all are known to glitch;
and many known glitching pulsars listed in the standard public catalogues~\cite{EspinozaGlitchCatalogue,ATNFGlitchDatabase}
are unfortunately not within the \aligo sensitivity band.
The search presented in this paper is intended as a pilot instance of this new type of transient analysis,
and the Crab and Vela pulsars were natural choices as the highest-priority targets.

To this end,
we for the first time apply the transient \Fstat method introduced
by Prix, Giampanis \& Messenger~\cite{Prix:2011qv}
to actual LIGO data.
It focuses on long-duration
(hours to months)
transient \gw signals that are quasi-monochromatic,
i.e. narrowly localized in frequency
(near twice the pulsar's rotation frequency)
at each point during the observation time
and slowly evolving in both frequency and amplitude.
The data is split into \sft[s]
and then,
for a bank of templates
from a simple frequency-evolution and transient-window model,
a maximum-likelihood matched-filter statistic is computed.
The method does not assume a particular glitch model,
but only that the \gw emission resembles
these simple phenomenological templates.

We present the data set used for this study in Sec.~\ref{sec:data},
general aspects of the signal model and \Fstat method in Sec.~\ref{sec:method}
and the specific search setup used in Sec.~\ref{sec:setup}.
Sec.~\ref{sec:results} summarizes the search results
and provides \ul[s] on \gw strain under the targeted signal model.
In Sec.~\ref{sec:discussion} we discuss their implications
and an outlook for future applications and refinements of this approach.

\section{\label{sec:data}Data used}

We use data from the O2 run
of the two \aligo detectors~\cite{TheLIGOScientific:2014jea}
in Hanford (H1) and Livingston (L1).
We start from the standard set~\cite{O2SFTs} of
1800\,s long
Tukey-windowed non-overlapping
\sft[s] used in previous \cw searches,
e.g.~\cite{Pisarski:2019vxw}.
(See Sec. IV.C.1 of~\cite{Abbott:2003yq} for the general construction of \sft[s].)
These were produced from the C02 version of calibrated strain data~\cite{Cahillane:2017vkb,Viets:2017yvy,Kissel:2018cal}
with some subtraction of known noise sources~\cite{Driggers:2018gii,Davis:2018yrz}.
As detailed in the next section,
we use about four months of data for each glitch search
(using YYYYMMDD notation for dates in the following):
20161211--20170411 for Vela
and
20170327--20170726 for the Crab,
covering
(3597,\,3156)
and
(2793,\,3091)
SFTs from (H1,L1) respectively.
This corresponds to effective duty factors of
(62\%,\,54\%)
and
(48\%,\,53\%).
Coincidence on a per-SFT level is not required for this analysis.

Two periods of O2, of about one month each,
of interest to this search
have been marked as spectrally contaminated in a single detector~\cite{O2SFTs}
and are not included in the standard \sft[s]:
L1 data before 20170104
and H1 data from 20170315--20170418.
We have investigated these data ranges in more detail
and for the two narrow frequency bands we consider here
(around twice the Vela and Crab rotation frequencies),
we have found that
there are not prohibitively many strong additional contaminations in these months
which are not also present in adjacent data.
Hence,
we have generated additional SFTs for these ranges
from frame data publicly available on GWOSC~\cite{gwosc:O2},
using the script lalapps\_MakeSFTDAG~\cite{lalsuite}
with exactly the same settings as for the standard set.

Some additional cleaning of narrow disturbances (`lines') was necessary.
A detailed list of \aligo O2 lines of relevance to \cw searches,
with known instrumental causes,
has been provided by~\cite{Covas:2018oik};
but since we are interested only in two narrow frequency bands
and in particular in transient disturbances,
we have also performed a separate,
simpler but more targeted
line identification exercise.
Details are given in
appendix~\ref{sec:cleaning}.

\section{\label{sec:method}Signal model and search method}

Following~\cite{Prix:2011qv},
our signal model is a slowly-varying quasi-monochromatic transient
\begin{equation}
 h(t,\Dop,\Amp,\TP) = \win(t,t_0,\tau)\,h(t,\Dop,\Amp) \,,
\end{equation}
corresponding to a classic \cw signal $h(t,\Dop,\Amp)$~\cite{Jaranowski:1998qm}
with an additional window function $\win(t;t_0,\tau)$
where the transient parameters $\TP$ consist of
the window shape,
signal start time $t_0$
and a duration parameter $\tau$.
The \cw part depends on
a set of phase evolution parameters
\mbox{$\Dop=\{\alpha,\delta,\fgw,\fdot,\fddot,\dots\}$}
(sky position,
frequency,
and frequency derivatives or `spindowns')
and on four amplitude parameters
\mbox{$\Amp=\{h_0,\cos\iota,\psi,\phi_0\}$},
which our detection statistic
(the \Fstat first introduced in~\cite{Jaranowski:1998qm})
analytically maximizes over.
($h_0$: dimensionless amplitude,
$\iota$ and $\psi$: orientation and polarization of the source,
$\phi_0$: \gw phase at reference time $\Tref$.)
The signal amplitude for a deformed \ns
at distance $d$,
emitting \gw[s] at \mbox{$\fgw=2\fspin$}
in the dominant \mbox{$l=m=2$} mode
from a quadrupolar ellipticity $\epsilon$,
with principal moment of inertia $\Izz$,
is given by
\begin{equation}
 h_0 = \frac{4\pi^2 G}{c^4} \frac{\epsilon\,\Izz}{d} \fgw^2 \,.
\end{equation}
Hence, the simplest interpretation of our transient signal model
would be a temporary increase in the quadrupolar deformation after the glitch,
with the product $\epsilon\,I$ falling off again
as determined by the transient window function
with timescale $\tau$.
This would be an intuitive behaviour within
the `starquake' model of pulsar glitches~\cite{Ruderman:1969nat,Smoluchowski:1970zz,Middleditch:2006ky},
where the crust is violently deformed.
In the more popular class of glitch models based on
a rotation lag between the bulk of the \ns and an interior superfluid component,
the interpretation becomes more complicated,
and could involve processes such as
crustal heating~\cite{VanRiper1991:xra},
non-axisymmetric oscillations~\cite{Bennett:2010tm}
or post-glitch excitation of Ekman flows~\cite{vanEysden:2008pd,Bennett:2010tm,Singh:2016ilt}.
These options are reviewed in more detail in ~\cite{Prix:2011qv,Haskell:2015jra}.
For our search, the mechanism does not initially matter
as long as a bank of templates from the simple signal model
provides a sufficient fit to the real signal.
The connection to physical models can be made later,
based on the signal durations and amplitudes observed,
or from the obtained \ul[s].

\begin{table*}
 \caption{
  \label{tbl:targets}
  Search targets and their key parameters.
  Much higher-precision sky location $(\alpha,\delta)$
  and frequency evolution ($f,\fdot,\fddot$)
  from the radio ephemerides
  are used in the search,
  with a factor 2 to convert from rotation to \gw emission rates,
  and ranges
  of 0.1\,Hz in $f$
  and
  $\approx10^{-13}$\,Hz\,s$^{-1}$ in $\fdot$
  are explored around the nominal values.
  For the Vela pulsar,
  all parameters are referenced to MJD 58000,
  while for the Crab pulsar
  the position epoch is MJD $\approx53254$
  and the frequency epoch is MJD $\approx57185.127$.
  The glitch times $\Tgl$ (in GPS seconds)
  and glitch sizes $\Delta f/f$
  are taken from~\cite{EspinozaGlitchCatalogue}
 }
 \begin{ruledtabular}
  \begin{tabular}{ccccccccc}
   target          & $d$ [pc]                                   & $\alpha$ [rad] & $\delta$ [rad]     & $f$ [Hz] & $\fdot$ [Hz\,s$^{-1}$] & $\fddot$ [Hz\,s$^{-2}$] & $\Tgl$ [s] & $\Delta f/f$ \\
   Vela (\velapsr) &  287~\cite{Dodson:2003ai}                  & 2.2486         & -0.7885            & 22.3722  & $-3.12\cdot10^{-11}$   & $1.16\cdot10^{-19}$     & 1165577920 & $1.43\cdot10^{-6}$ \\
   Crab (\crabpsr) & 2000~\cite{Trimble:1973pasp,Kaplan:2008qm} & 1.4597         & \hphantom{-}0.3842 & 59.3295  & $-7.39\cdot10^{-10}$   & $1.72\cdot10^{-20}$     & 1174687506 & $2.14\cdot10^{-9}$ \\
  \end{tabular}
 \end{ruledtabular}
\end{table*}

For each target pulsar, $\alpha$ and $\delta$ are fixed.
We place a grid in $(\fgw,\fdot)$ space with fixed spacings.
The search method, implemented in the program
lalapps\_ComputeFstatistic\_v2~\cite{lalsuite},
then loops over this grid,
and for each \mbox{$\lambda=(\fgw,\fdot)$} pair
it computes the transient $\F$-statistic map
\begin{equation}
 \label{eq:Fmn}
 \Fmn(\Dop) = \F(\Dop,t_{0\,m},\tau_n) 
\end{equation}
by computing partial sums,
corresponding to all $\{t_{0\,m},\tau_n\}$ combinations,
of the per-\sft `atomic' ingredients~\cite{Prix:2011qv,Prix:cfsv2} of the \Fstat.
This corresponds to demodulating~\cite{Williams:1999nt} the detector data
over the whole observation time
once for each $\Dop$ template,
taking into account the time-varying detector response,
and then simple arithmetic operations for the transient aspect.

In this search, we use a simple `rectangular window',
i.e. \mbox{$h_0=\mathrm{const.}$}
for \mbox{$t \in [t_0,t_0+\tau]$}
and \mbox{$h_0=0$} outside.
The maximum loss compared to more realistic
exponential-decay windows has already been estimated as acceptable in \cite{Prix:2011qv}.
A search with exponential windows would also be computationally much more costly;
see~\cite{Keitel:2018pxz} for a detailed discussion.
Also see Sec.~\ref{sec:discussion} and appendix~\ref{sec:exp} for more details on this point.

To obtain detection candidates,
we are interested in peaks over the full
\mbox{$(\Dop,\TP)=(\fgw,\fdot,t_0,\tau)$} space.
To reduce the number of strong \Fstat outliers
due to single-detector noise disturbances,
we also apply the line-robust statistic $\BSGL$
from~\cite{Keitel:2013wga}
at every point,
and for each $\Dop$ we only store the values of
$\max\limits_{t_0,\tau}\Fmn(\Dop)$
over the subset where $\BSGL(\Dop,t_0,\tau)$
is above some threshold.
In the end,
candidates are identified from the $\F$-ordered results,
as the \Fstat is directly related to the \snr
and its analytically known distribution
will allow us to set a detection threshold without additional pure-noise Monte Carlos.
$\BSGL$ is thus used here simply as an intermediate veto step,
not as a full replacement detection statistic.

\section{\label{sec:setup}Search setup and parameter space covered}

The key parameters of the two target pulsars and their glitches during O2
are summarized in Table~\ref{tbl:targets}.
For each glitch, we search for transients starting in a \mbox{$\Delta t_0=1$\,day} window
centred around the nominal glitch epoch
and with durations $\tau$ up to 121\,days.
Hence we search a data set of $\Tobs\approx4$\,months,
chosen for two reasons:

(i) With the whole of O2 lasting about
9
months,
for much longer $\tau$ we would no longer get sufficient benefits
from the transient $\F$-statistic over the results
of full-O2 \cw analyses~\cite{Abbott:2019ztc,Abbott:2019bed}.
From Eq.~(62) in \cite{Prix:2011qv}, the mismatch (relative loss in squared \snr)
for observing a signal of true length $\tau_s$ with a rectangular template window of length $\tau$ is
\mbox{$m\approx1-(\tau-\tau_s)^2/(\tau\tau_s)$}.
So for a maximum \mbox{$\tau_s=4$\,months}
and a full-O2 \cw search's \mbox{$\tau=9$\,months},
we expect
\mbox{$m\approx30$\%}
corresponding to still about a factor of 2 gain in \snr
for a transient with our maximum $\tau$ compared
to the full-O2 CW search.
For longer $\tau_s$,
the gain
would be correspondingly smaller.

(ii) $\Tobs\approx4$\,months also matches the duration
of the first \aligo observing run (O1),
so that,
at the longest $\tau$,
our search becomes comparable in setup to the O1 narrow-band search~\cite{Abbott:2017cvf}
and we can draw some direct comparisons in the following.

Similar to \cite{Abbott:2017cvf,Abbott:2019bed},
we allow for some mismatch between the true \gw frequency $\fgw$
and its nominal value
(twice the radio-observed $\fspin$),
constructing a rectangular search grid in $(\fgw,\fdot)$ space.
We choose resolutions in \gw frequency and spin-down of
\mbox{$d\fgw  = 1/\Tobs   \approx 9.57\times10^{-8}$\,Hz}
and
\mbox{$d\fdot = 1/\Tobs^2 \approx 9.15\times10^{-15}$\,Hz\,s$^{-1}$},
and cover a frequency range of
0.1\,Hz
and a spindown range of
\mbox{$11d\fdot\approx1.01\times10^{-13}$\,Hz\,s$^{-1}$}
both centred on a point $(\fgw,\fdot)$ corresponding to
twice the values from the pulsar's radio ephemerides.

The ephemerides were obtained from observations at
the University of Tasmania's Mount Pleasant Radio Observatory
for Vela
and at Jodrell Bank (UK)
for the Crab,
using the TEMPO2 software~\cite{Hobbs:2006cd}.
Both were originally fitted
with the goal of minimal residuals
over the whole respective data ranges of the \cw searches in~\cite{Abbott:2019ztc}.
Vela, in addition to glitch recovery,
has a lot of timing noise~\cite{Hobbs2006:ptn,Ashton:2014qya}
and micro-glitches~\cite{Cordes1988:dkp};
hence, to minimize overall residuals,
the post-glitch ephemeris for O2
was fitted down to the twelfth derivative
(without an exponential recovery term).
Thus,
and due to the intrinsic rapid evolution of the frequency and its derivatives
after a glitch,
the $\fddot$ in Table~\ref{tbl:targets} should not be compared directly
with the long-time value reported by~\cite{Lyne:1996nat}.
Similarly, the Crab pulsar also has significant timing noise and glitch recovery
complicating its spin-down~\cite{Lyne:2014qqa}.
For its O2 ephemeris,
no explicit glitch model was used,
but again twelve spin-down derivatives were included
to minimize overall residuals.

\begin{figure*}
 \includegraphics{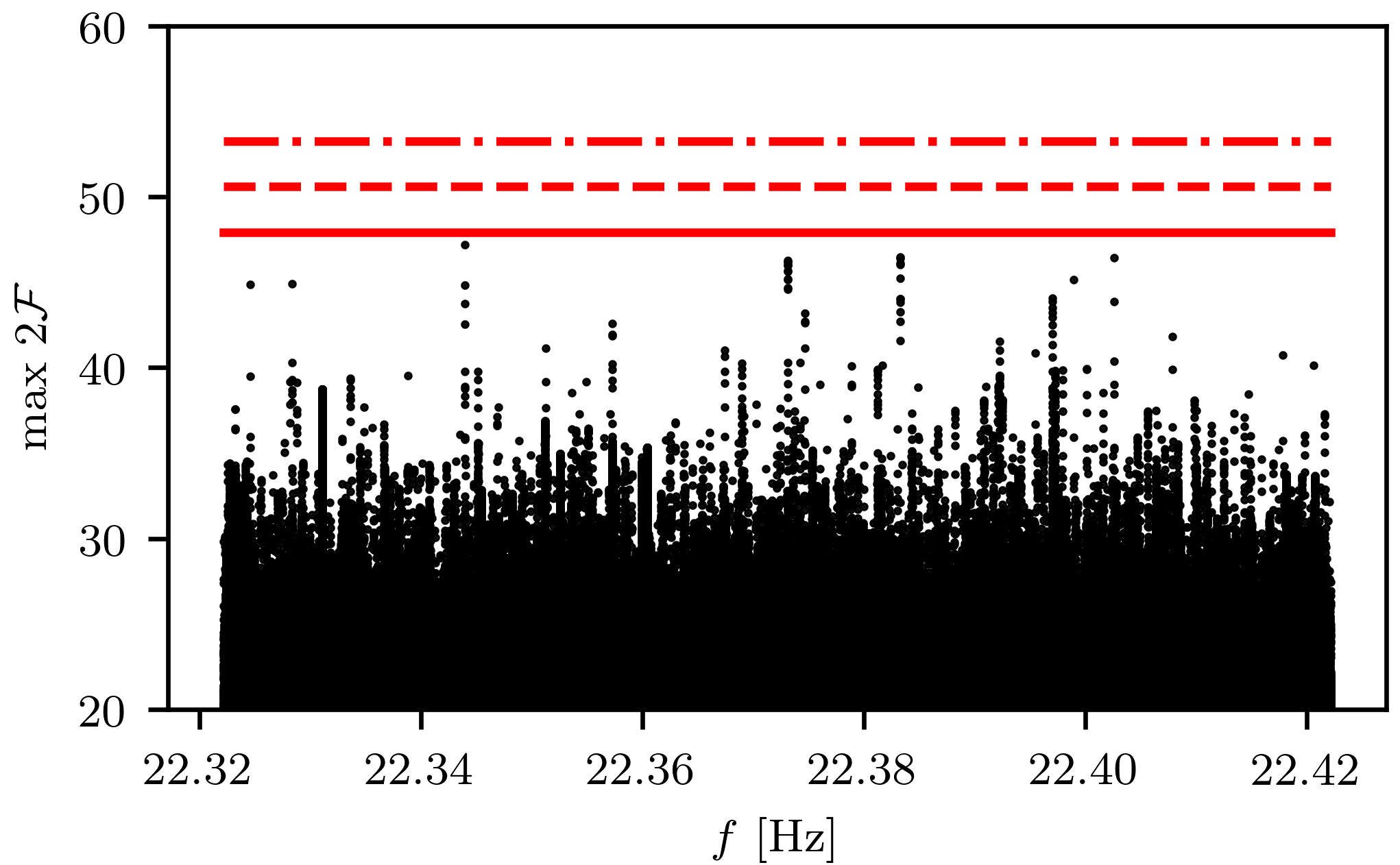}
 \includegraphics{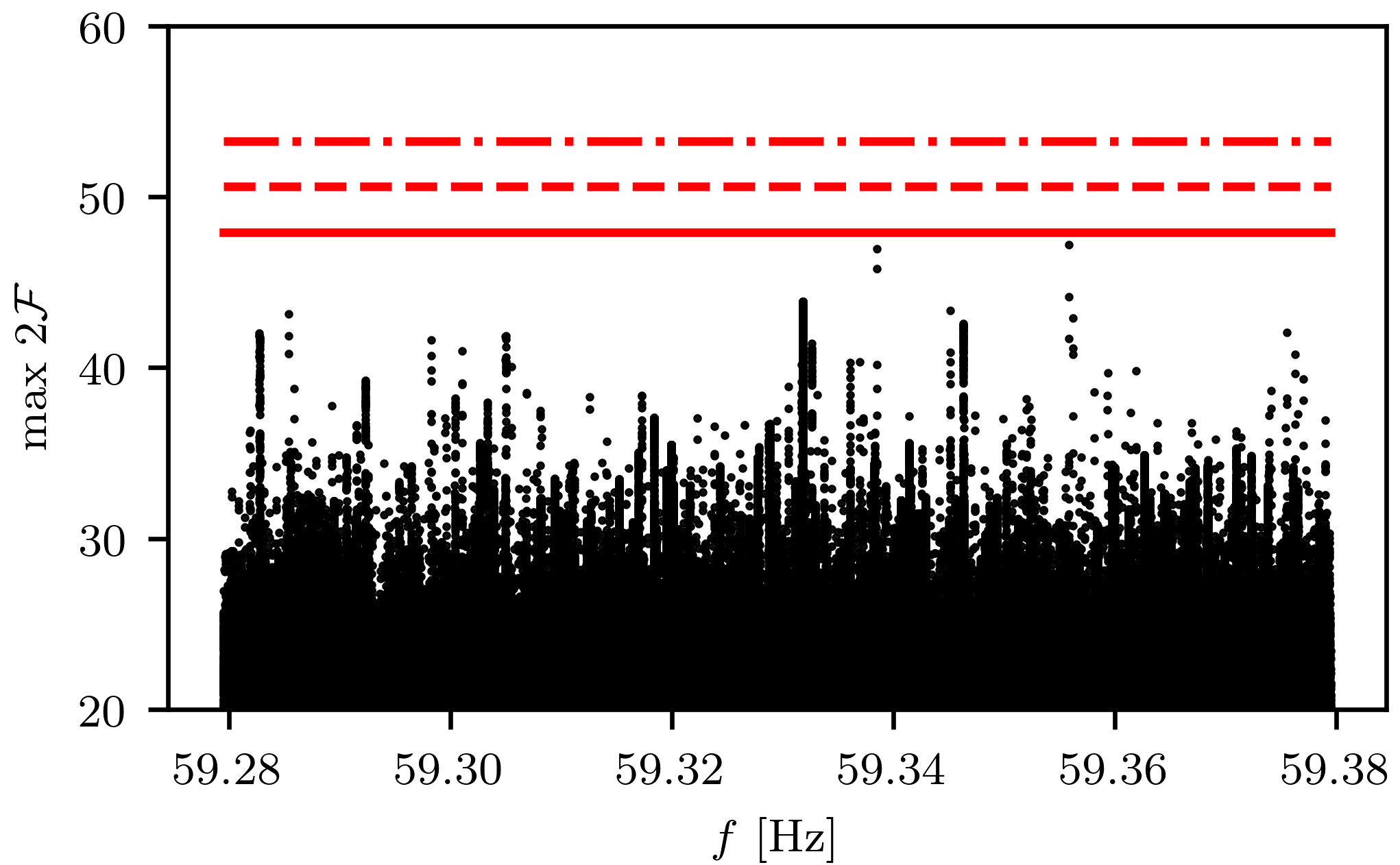}
 \caption{\label{fig:freq-max2F}
  Search results for
  transients following
  the 20161212 Vela glitch (left panel)
  and 20170327 Crab glitch (right panel)
  with the detection statistic $\max2\F(f,\fdot)$
  maximized over transient parameters $\{t_0,\tau\}$ and projected
  onto the $f$ axis.
  Horizontal lines correspond to expected maximum outliers
  (and also +1 and +2 standard deviations)
  for an estimated effective template count
  (per target)
  of $5.75\cdot10^8$.
 }
\end{figure*}

For both targets,
we use the $(f,\fdot$) grid as described above,
a single fixed value for $\fddot$,
and set higher-order terms to zero,
which is valid in the sense that we are
accumulating less than a bin of mismatch in $f$ over our $\Tobs$
from the second derivative and higher.
In addition, while both ephemerides
have reference times far from the glitches,
the ranges covered in $f$ and $\fdot$ mean that either
we resolve the glitch step with multiple templates (Vela),
or the glitch step is itself smaller than our search resolution (Crab),
so that extrapolation to around the glitch epoch
is safe for the purpose of this search.

The overall frequency evolution template count is
\mbox{$\Ntemp = N_\fgw N_\fdot \approx 1.15\times10^7$}.
Comparing with \cite{Abbott:2017cvf},
these choices mean we cover slightly wider ranges in $\fgw$ and $\fdot$ for Vela
while for the Crab we cover the same $\fgw$ and a narrower $\fdot$ range by a factor 15.

For each $\{\fgw,\fdot\}$ parameter space point,
transient parameters of
\mbox{$t_0\in[\Tgl-0.5\,\days,\Tgl+0.5\,\days]$}
and \mbox{$\tau\in[0.5\,\days,\Tobs]$}
are analyzed with a resolution
\mbox{$d t_0=d\tau=\Tsft=1800$\,s},
yielding
\mbox{$\Ntrans=N_{t_0}N_\tau=48\times5784\approx2.8\times10^5$} grid points.

The lower limit of \mbox{$\tau\geq0.5\,\days$} is chosen
empirically to avoid spurious outliers
when statistics fluctuate too much over a small number of \sft[s].
In principle,
the transient-\Fstat method can be extended to much shorter $\tau$
but this would require additional copies of the data set with shorter $\Tsft$.
This could be done in the future,
or a more adaptive multi-timescale approach could be pursued;
but for this pilot search we limit ourselves to the standard data set with \mbox{$\Tsft=1800$\,s}
and correspondingly do not explore very short transients.

The line-robust $\BSGL$ statistic we use as a veto
has a free tuning parameter $\Ftho$
determining how strongly it is allowed to deviate from the standard \Fstat~\cite{Keitel:2013wga}.
Since in this search we are dealing with narrow frequency bands
which are already known to contain some disturbances,
we choose a relatively low value
\mbox{$\Ftho=10$}
corresponding to a statistic that would be slightly suboptimal in purely Gaussian noise
but is stricter in suppressing lines.
We then set a rather lenient threshold of
\mbox{$\BSGL>-10$}
to only cut out very strong single-detector artifacts
that might have passed our line cleaning procedure
(see appendix~\ref{sec:cleaning}),
without severely affecting the distribution of \Fstat values.

\section{\label{sec:results}Results}

\subsection{\label{sec:results-search}Search results: no significant candidates}

Full results for the $\max2\F(f,\fdot)$ statistic
(maximized over $t_0$ and $\tau$),
projected onto the frequency axis,
are shown
in Fig.~\ref{fig:freq-max2F} for both
the 20161212 Vela glitch
and the 20170327 Crab glitch.

To determine whether the loudest of these per-template results
constitute promising detection candidates,
we can consider the well-known~\citep[e.g.][]{Aasi:2013jya}
statistical properties of the $\F$-statistic:
as $2\F$ in pure Gaussian noise
follows a $\chi^2_4$ distribution
(with 4 degrees of freedom),
the loudest value $2\F^*$ from $N$ independent trials
is distributed as
\begin{equation}
  \label{eq:max2Fpdf}
  p(2\F^*;N) = N \, \chi^2_4(2\F^*) \, \chi^2_4(2F<2\F^*)^{N-1} \,,
\end{equation}
where $\chi^2_4(2\F^*)$ is the probability distribution function evaluated at $2\F^*$
and $\chi^2_4(2F<2\F^*)$ is the cumulative distribution function integrated up to $2\F^*$.

For this search, we need to consider the loudest overall outlier
\mbox{$2\F^*=\max_{f,\fdot}(\max_{t_0,\tau}2\F)$}
for each target.
Since the $(t_0,\tau$) templates at each $(f,\fdot)$ point re-use the same data atoms many times,
the total effective number of templates is much lower than $\Ntemp\cdot\Ntrans$.
We find acceptable fits to the $\max2F$ distribution (over all $(f,\fdot)$) 
for \mbox{$\Ntranseff\approx50$},
and then use a total \mbox{$\Neff\approx50\cdot\Ntemp$}
to obtain (numerically from Eq.~\ref{eq:max2Fpdf})
an expectation value
\mbox{$E[2\F^*]\approx48$}
with standard deviation
\mbox{$\sigma[2\F^*]\approx2.7$}.

\begin{figure*}
 \includegraphics{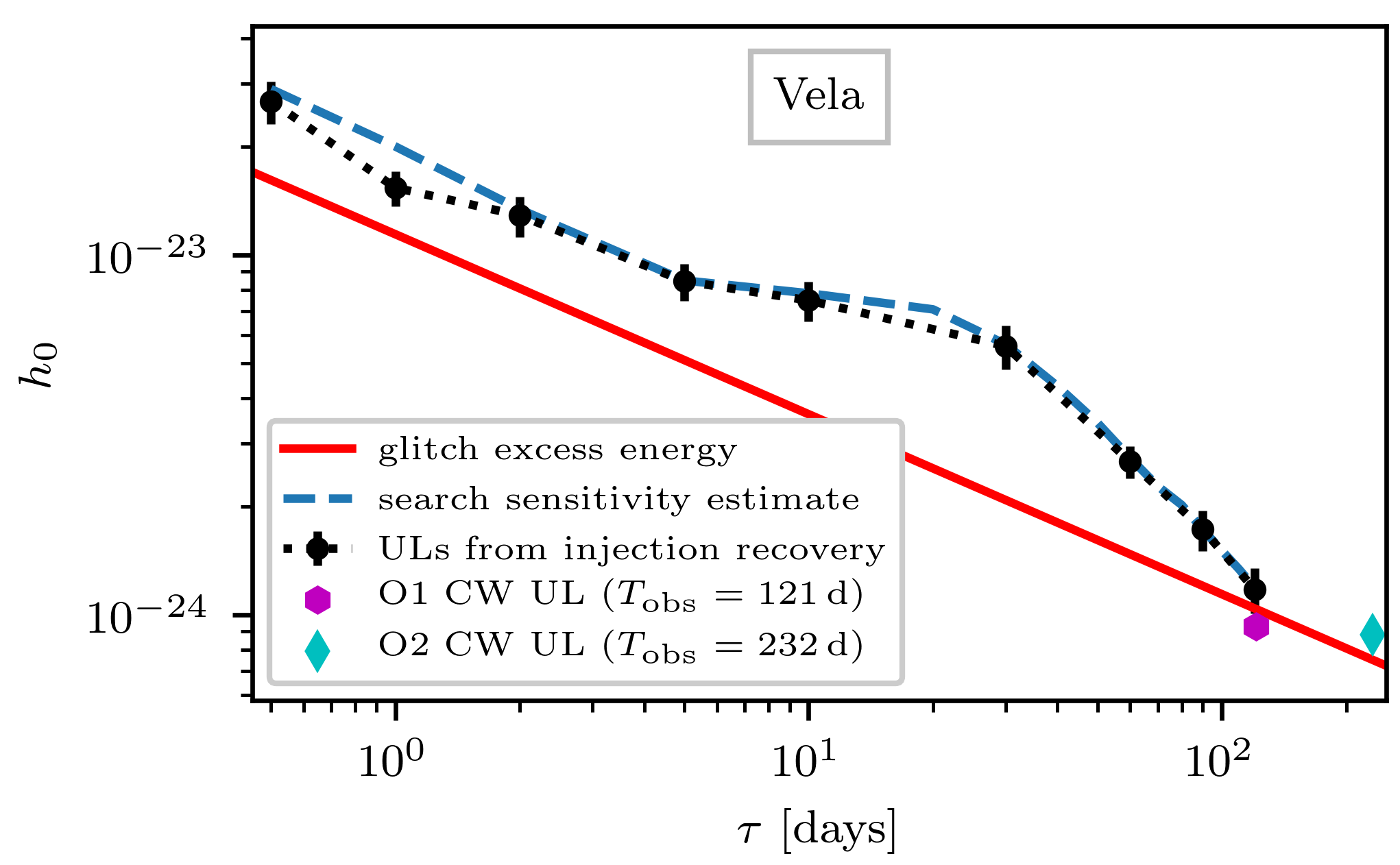}
 \includegraphics{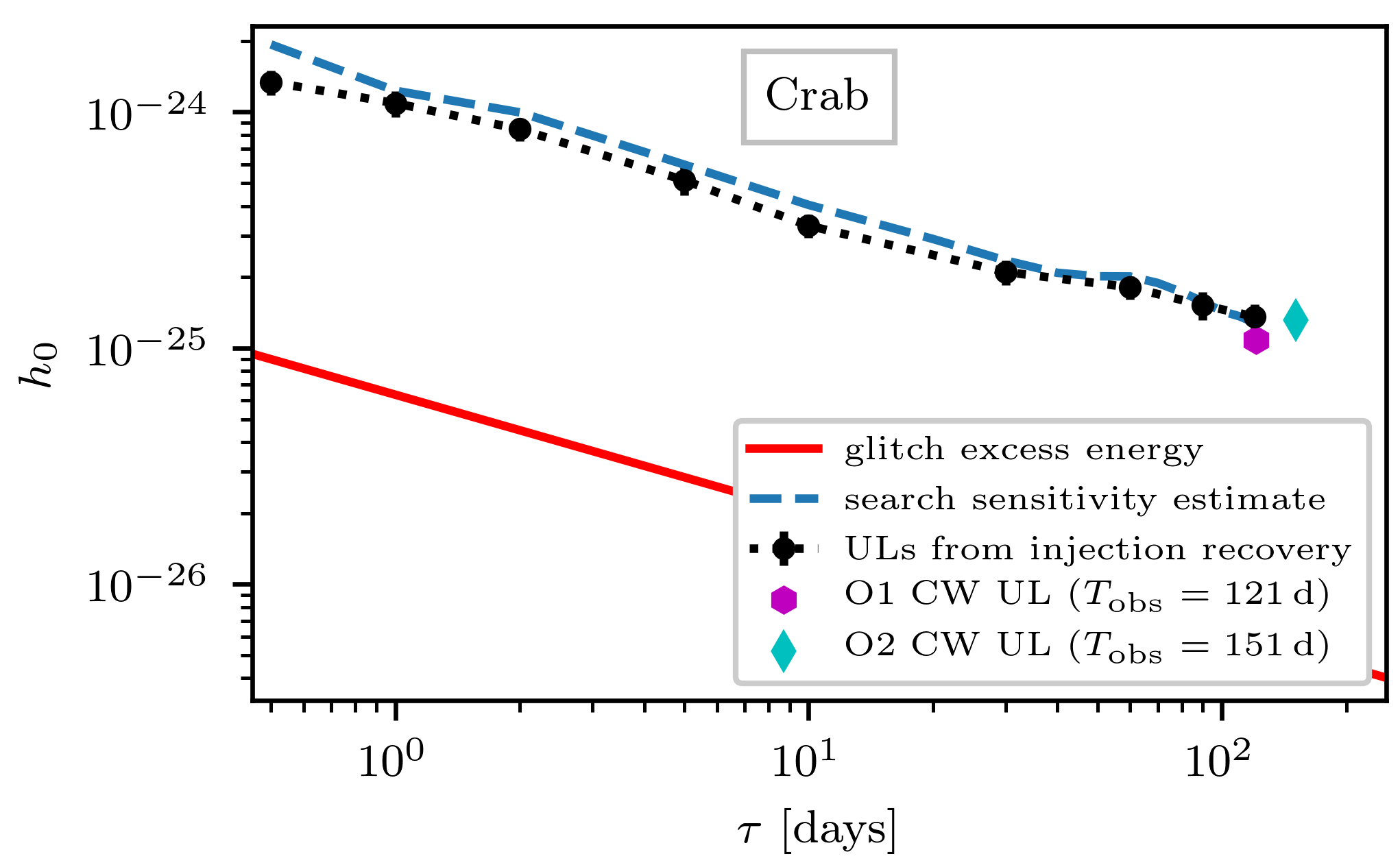}
 \caption{\label{fig:uls-tau}
  \ul[s] on \gw strain for transients following
  the 20161212 Vela glitch (left panel)
  and 20170327 Crab glitch (right panel)
  as a function of the duration $\tau$ of rectangular-windowed signals.
  The black points and dotted lines
  represent our \ul[s] at 90\% confidence
  from simulated signal injections
  randomized over all other parameters.
  Error bars are estimated from a sigmoid fit to $\pdet(h_0)$ curves,
  with a small contribution from calibration uncertainty
  (see appendix~\ref{sec:ulmethod}).
  For comparison, we also show
  the glitch excess energy indirect \ul[s] from Eq.~\ref{eq:energyUL}
  (solid red lines),
  a semi-analytic sensitivity estimate for the search
  (dashed blue lines)
  and the \ul[s] for persistent \cw[s] previously obtained in
  \cite{Abbott:2017cvf}
  (magenta hexagons,
  full O1 data)
  and \cite{Abbott:2019bed}
  (cyan diamonds,
  O2 data starting 20170104 for Vela
  and after 20170327 for the Crab).
 }
\end{figure*}

The loudest candidate from the Vela search has
\mbox{$2\F^*\approx47.19$},
and from the Crab search the loudest outlier is at
\mbox{$2\F^*\approx47.18$}.
These do not even cross a nominal
\mbox{$E[2\F^*]\approx48$} threshold,
which might be considered a low-level criterion for further follow-up study.
And even though the precise threshold to put could be shifted somewhat
when revisiting the assumptions in effective template counting,
any truly promising candidate would need to lie a few standard deviations above expectation.
For comparison,
levels of
$E$, $E+\sigma$ and $E+2\sigma$
are indicated by horizontal lines
in Fig.~\ref{fig:freq-max2F}.

Hence, while there is some evident substructure in the search results
that could be further investigated by follow-ups
with different $\Tsft$,
gridless MCMC~\cite{Ashton:2018ure}
or through more detailed data quality studies,
we conclude that the loudest candidates from both targets
are so weak that no such effort is warranted at this point.
We proceed next to set \ul[s]
on \gw strain from our targets
based on the absence of a detection.

\subsection{\label{sec:search-uls}Upper Limits}

To obtain \ul[s] on the emitted strain from/after the two targeted glitches,
still under the simplified assumption of a `rectangular' transient,
we perform software injections of simulated signals into the same data sets
as used for the original searches,
and determine the required scale of $h_0$
at which 90\% of signals are recovered above
the nominal \mbox{$E[2\F^*]\approx48$} threshold.
Physically,
for a fixed overall energy budget of the glitch
we would expect lower $h_0$ for longer $\tau$.
At the same time,
our search sensitivity also improves roughly with $\sqrt{\tau}$.
Hence,
we present \ul[s] on $h_0$ as a function of $\tau$,
randomizing over all other parameters.
Details on the procedure are given in appendix~\ref{sec:ulmethod},
and the results
are shown
in Fig.~\ref{fig:uls-tau}.

In the same figures, we also compare the measured \ul[s]
against a sensitivity estimate for our search
based on the \texttt{octapps}~\cite{Wette:2018rsw} \cw sensitivity calculator
(see~\cite{Dreissigacker:2018afk} for a detailed description).
To obtain this curve,
we consider \cw-like searches over a grid in $\Tobs=\tau$
and feed the estimation code
with cumulative duty factors and harmonic-mean averages of the detector \psd over each of these durations,
as well as an estimated average mismatch of 0.2
for our $(f,\fdot)$ grid
and a $2\F$ threshold of 48 as above.
The agreement is remarkably close.

We also find that at the longest signal duration probed
(\mbox{$\tau=120$\,d})
our \ul[s]
approach those from the LIGO-Virgo narrow-band searches for \cw[s]
from the same targets
(see Table VI of~\cite{Abbott:2017cvf} and Table IV of ~\cite{Abbott:2019bed}).

\section{\label{sec:discussion}Discussion}

In the absence of significant \gw detection candidates
after the two glitches in the Vela and Crab pulsars during the \aligo O2 run,
we have set \ul[s] on the emitted \gw strain as a function of signal duration $\tau$.
To check how physically constraining these \ul[s] are,
we consider an indirect transient energy \ul in analogy to the well-known spin-down limit from the \cw case
(see e.g. Sec.~2.3 of~\cite{Prix:2009oha}).
Here we briefly summarize the derivation of this indirect \ul from Sec.~II.C of~\cite{Prix:2011qv}.

According to the basic two-fluid model of \ns glitches
(see e.g.~\cite{Lyne:2000sta}),
in an intra-glitch period the bulk of the \ns gradually spins down,
while an interior superfluid component retains most of its initial angular momentum.
Then assuming
(i) a glitch transfers the whole angular momentum difference
previously built up between the two components,
(ii) the moments of inertia
$\Isuper$ (superfluid)
and $\Inormal$ (bulk)
do not change during the glitch,
and (iii) $\Isuper \ll \Inormal$,
then the excess superfluid energy liberated in this transfer is
\mbox{$\Egl \approx 4\pi^2 \, \Isuper \, f \lag$},
where $\lag$ is the lag
(built-up difference in $f$) between the two components
right before the glitch
($\Delta \nu$ in the notation of~\cite{Prix:2011qv}).

If we further assume that
(iv) all of $\Egl$ is emitted in \gw[s],
and rewrite in terms of the observed relative frequency change $\Delta f/f$ at the glitch
and total moment of inertia $\Izz\approx\Inormal$,
then the total emitted \gw energy is independent of the signal duration $\tau$,
while the corresponding \gw strain as a function of $\tau$ is
\begin{equation}
 \label{eq:energyUL}
 h_0 = \frac{1}{d} \sqrt{\frac{5G}{2c^3} \frac{\Izz}{\tau} \frac{\Delta f}{f}} \,.
\end{equation}
As per~\cite{Prix:2011qv},
qualitatively similar \ul[s] still hold
for alternative glitch models
such as crust-cracking starquakes~\cite{Middleditch:2006ky},
where instead the moment of inertia of the crust would change at the glitch.

This indirect \ul is shown for comparison with our empirical \ul[s]
in Fig.~\ref{fig:uls-tau}.
For this,
we assume a fiducial value of
\mbox{$\Izz=10^{38}$\,kg\,m$^2$}
and distances of
287\,pc for the Vela pulsar~\cite{Dodson:2003ai}
and 2\,kpc for the Crab pulsar~\cite{Trimble:1973pasp,Kaplan:2008qm}.

With a frequency change of
\mbox{$\Delta f/f = 1.431\cdot10^{-6}$},
the Vela glitch was much larger
than the Crab glitch with
\mbox{$\Delta f/f = 2.14\cdot10^{-9}$}
(both values from~\cite{EspinozaGlitchCatalogue};
\cite{Ashton:2019hqd} recently suggested a larger initial frequency overshoot in the Vela glitch,
which however has already decayed at the timescales probed in this search).
Hence, though \aligo sensitivity
is better at the higher of the two target frequencies,
for the Crab glitch our O2 search was still far away from the indirect energy \ul;
while for Vela we got very close to beating it
at the shortest and longest $\tau$.
Overall, the search sensitivity was mostly limited
by the significant time variation
of detector sensitivities and duty factors.
For example,
the slow improvement in Vela \ul[s] at intermediate durations
(10--30 days)
is largely due to the winter holiday break in O2 observing
which made cumulative duty factors drop over this period.

Since the search presented in this paper
is the first practical application of the
transient-\Fstat method from~\cite{Prix:2011qv}
to \gw detector data,
we have made several choices to simplify the search setup
and post-processing,
but which can be improved over in future applications.
Notably, the simple rectangular transient window allows for efficient computation,
but while it does also recover most of the \snr for different signal shapes
(see~\cite{Prix:2011qv} and appendix~\ref{sec:exp}),
a more general and sensitive analysis will be possible when including
e.g. exponentially decaying window functions.
(A natural choice considering
the observed exponential recoveries in pulsar frequency after most glitches,
see e.g.~\cite{Haskell:2013goa}.)
Since exponential templates are much more costly,
using \gpu[s] for the search would be highly beneficial~\cite{Keitel:2018pxz}.

The search setup can be improved in several other ways,
including multi-timescale approaches:
i.e. the use of input data at several different \sft durations
to cover shorter transients;
and varying the metric-based~\cite{Prix:2006wm} frequency and spin-down resolutions
as a function of $\tau$.

To deal with outliers from either detector noise features or actual \gw signals,
MCMC-based follow-up~\cite{Ashton:2018ure} is a promising technique.
If necessary, the detector noise can also be studied in much more detail
than was required for this pilot search,
e.g. using correlations with auxilliary channels to veto instrumental lines~\cite{Covas:2018oik}.
We also used only an ad-hoc veto version of the line-robust statistic from~\cite{Keitel:2013wga},
while a semi-coherent transient-aware version~\cite{Keitel:2015ova}
or a customized coherent version would offer
the potential for more robust suppression of single-detector instrumental artifacts.

In addition to these improvements in the analysis method,
the improved sensitivity and increased number of detectors
in O3 (which has started in April 2019) and beyond~\cite{Aasi:2013wya}
will be the strongest driver in bringing search results for future pulsar glitches
into the physically constraining regime.
Since the search method is computationally cheap
(already on regular CPUs for rectangular windows,
and when using \gpu[s]~\cite{Keitel:2018pxz} this stays true also for exponential windows),
with some more automation of the analysis pipeline
it should be possible to target not only the highest-value objects such as the Vela and the Crab,
but more of the large population of glitching pulsars~\cite{Fuentes:2017bjx}.

\begin{acknowledgments}
We thank members of the LIGO-Virgo Continuous Wave working group for many fruitful discussions;
Reinhard Prix and Chris Messenger for initial advice on the transient-\Fstat method;
Pep Covas Vidal,
Evan Goetz,
Ansel Neunzert
and the rest of the LIGO detector characterization group
for their invaluable work on data quality studies;
Greg Ashton for detailed comments on the manuscript;
and Paul Hopkins and Stuart Anderson for technical support.
For part of this project,
DK was funded under the EU Horizon2020 framework
through the Marie Sk\l{}odowska-Curie grant agreement 704094 GRANITE.
GW and MP are funded through the
UK Science \& Technology Facilities Council (STFC) grant ST/N005422/1.
CS was supported through the Arcadia University Study Abroad programme.
This research has made use of data obtained from the Gravitational Wave Open Science Center,
a service of LIGO Laboratory, the LIGO Scientific Collaboration and the Virgo Collaboration.
LIGO is funded by the U.S. National Science Foundation.
Virgo is funded by the French Centre National de Recherche Scientifique (CNRS),
the Italian Istituto Nazionale della Fisica Nucleare (INFN)
and the Dutch Nikhef,
with contributions by Polish and Hungarian institutes.
The authors are grateful for computational resources provided by
the LIGO Laboratory
and Cardiff University
and supported by
National Science Foundation Grants PHY-0757058 and PHY-0823459
and STFC grant ST/I006285/1.
This paper has been assigned document number \dcc.
\end{acknowledgments}

\appendix
\section{\label{sec:cleaning}Data cleaning}

To identify single-detector noise artifacts in the form of
narrow disturbances (spectral lines)
within the analysis bands for each pulsar,
we have considered a standard quantity for such data quality studies,
the normalized SFT power~\cite{Abbott:2003yq}:
\begin{equation}
  \label{eq:Psft}
  \Psft^X(f_k) = \frac{2}{\Nsft\,\Tsft}
               \frac{ \left| \widetilde{x}_\alpha^X(f_k)\right|^2 }{\SnXal(f_k)}\,.
\end{equation}
Here,
$\widetilde{x}_\alpha^X(f)$ is the data in bin $f_k$ of the $\alpha^{\mathrm{th}}$ SFT
for detector $X$
and $\SnXal(f_k)$ is a running-median noise \psd estimate.
This is computed for the bands in question
using the lalapps\_ComputePSD tool~\cite{lalsuite}.
We then identified as (possibly transient) single-detector line features
those frequency bins
(or small number of adjacent bins)
where at least 5 SFTs
(not necessarily consecutive)
crossed a threshold
\mbox{$\Psft^X(f_k)>10$}.
These are listed in Table~\ref{tbl:cleaned-lines}.
For safety,
we would not have vetoed any features common to both detectors --
as could be produced by an (extremely loud) astrophysical signal --
through this procedure.
No such coincident features were found in the two bands.

Comparing with the detailed detector characterization approach of line hunting in~\cite{Covas:2018oik},
we find that the three easily identified H1 artifacts match up with harmonics of known instrumental frequency combs.
On the other hand, the three L1 artifacts are all very short and limited to the first month of O2
when an improperly connected ethernet cable induced electronic crosstalk in the interferometer controls system,
a period which has not been used in the LIGO-Virgo flagship \cw searches~\cite{Abbott:2019ztc,Abbott:2019bed,Pisarski:2019vxw}.
These artifacts do not match any lines or combs listed in~\cite{Covas:2018oik}.
We have not investigated in any more detail whether these additional narrow and transient disturbances
are clearly correlated with auxilliary channels,
but since they are clearly limited to a single detector,
they can still be considered safe for removal from the input data.

Using this list,
we have cleaned the input \sft[s] of the affected detector
by replacing the listed bins with samples
drawn from a Gaussian distribution with variance matching the surrounding PSD estimate,
using lalapps\_SFTclean.

After this removal,
no outliers with  multi-detector \mbox{$\max2\F\geq48$}
or single-detector \mbox{$\max2\F\geq52$}
are found by the search.
So while some of the remaining substructures
visible in
Fig.~\ref{fig:freq-max2F}
are likely due to unidentified narrow instrumental disturbances,
possibly including the December 2016 L1 and March--April 2017 H1 issues,
none of these are strong enough to lead to significant outliers.

\begin{table}
 \caption{
  \label{tbl:cleaned-lines}
  Single-detector lines identified through outliers in the normalized SFT power
  and subsequently removed from the input \sft[s] for the \Fstat analyses.
  The more common cases where single bins in a single SFT exceed the threshold
  are not used for any cleaning.
  The width is given in terms of the number of bins $N_\mathrm{bins}$
  at resolution \mbox{$1/\Tsft=1/1800$\,s}.
  The duration is quoted as between the first and last \sft
  with \mbox{$\Psft^X(f_k)\geq10$},
  not necessarily meaning that the line is persistently visible during this range,
  and not excluding that it is still present at a weaker level before and after.
  For simplicity and safety,
  the full set of \sft[s] is cleaned at these frequencies,
  not just the listed duration.
  Where the frequency matches up with a harmonic of a known comb of disturbances,
  the corresponding comb spacing is listed.
 }
 \begin{ruledtabular}
  \begin{tabular}{ccccc}
   detector & $f_\mathrm{start}$ & $N_\mathrm{bins}$ & duration             & comb [Hz] \\
   H1       & 22.2222            & 3                 & 20161202--20170508   & 11.1111~\cite{Covas:2018oik} \\
   H1       & 22.2500            & 1                 & 20170204--20170418   & 0.9999862~\cite{Covas:2018oik} \\
   L1       & 22.3600            & 3                 & 20161214             & \textit{unidentified} \\
   L1       & 22.4156            & 3                 & 20161210--20170104   & 2.24154~\cite{Riles:alog29853,Effler:alog30655} \\
   L1       & 22.4706            & 1                 & 20161213             & \textit{unidentified} \\
   H1       & 59.2733            & 3                 & on/off through O2 & 0.9878881~\cite{Covas:2018oik} \\
  \end{tabular}
 \end{ruledtabular}
\end{table}

\section{\label{sec:ulmethod}Details on \ul procedure}

To obtain the \gw strain \ul[s] as a function of signal length $\tau$
as presented in Sec.~\ref{sec:search-uls},
we perform software injections of simulated signals
into the same input \sft[s] as used for the main search.
At each sample value of $\tau$,
we use the lalapps\_MakeFakeData\_v5 program~\cite{lalsuite}
to simulate a set of signals with varying $h_0$,
uniformly distributed over the whole search range in $\{f,\fdot,t_0\}$
and also randomized over the remaining amplitude parameters $\{\cos\iota,\psi,\phi_0\}$.
For each injection, a reduced parameter space covering
$10^{-4}$\,Hz in frequency (with the same $df$ as before)
and only 1 spin-down bin is re-analyzed.
We count an injection as recovered if it produces \mbox{$\max2\F\geq48$},
as a candidate above this nominal threshold
would have been considered for further follow-up if found in the main search.

For each $\tau$,
the result is then an efficiency curve
of detection probability $\pdet$ against injected $h_0$.
This can be fit with a sigmoid
\begin{equation}
 \pdet(h_0) = \frac{1}{1+\exp[-a(h_0-b)]}
\end{equation}
(using \texttt{scipy.curve\_fit})
and evaluated at \mbox{$\pdet=0.9$}
to estimate a $\hul$ \ul at 90\% confidence.
The error on $\hul$ is obtained from error propagation
of the uncertainties in the fit coefficients $(a,b)$.
We have run a relatively small set of \ul simulations:
50 injections each at 10--20 $h_0$ steps per $\tau$ value,
leading to $\sim10$\% uncertainty in $\hul$ as evaluated from the fit.
An additional contribution comes from calibration uncertainty in the measured strain at the detector~\cite{Cahillane:2017vkb,Viets:2017yvy}.
According to~\cite{Kissel:2018cal},
for the 20--100\,Hz band during O2
the amplitude uncertainties are 1.6\% for H1 and 3.9\% for L1.

\section{\label{sec:exp}Sensitivity compared with exponentially decaying transients}

In this search we have only considered the simplest model for quasi-monochromatic transients,
i.e. \cw-like signals turning on at some time $t_0$ and off again at $t_0+\tau$
with fixed amplitude,
hence referred to as rectangular-windowed signals.
The detection method as introduced by~\cite{Prix:2011qv}
however can also deal with more general transient window functions,
and as an explicit example
the implementations in LALSuite~\cite{lalsuite} and pyFstat~\cite{Keitel:2018pxz}
also include transients with exponentially decaying amplitude.
Since pulsar rotation frequencies after glitches
often show an exponential recovery profile~\cite{Haskell:2013goa},
exponential transient windows could be a more realistic option.
Here we briefly investigate
the effect of
trying to recover exponentially-decaying signals with rectangular search windows.

\begin{figure}[t!]
 \includegraphics{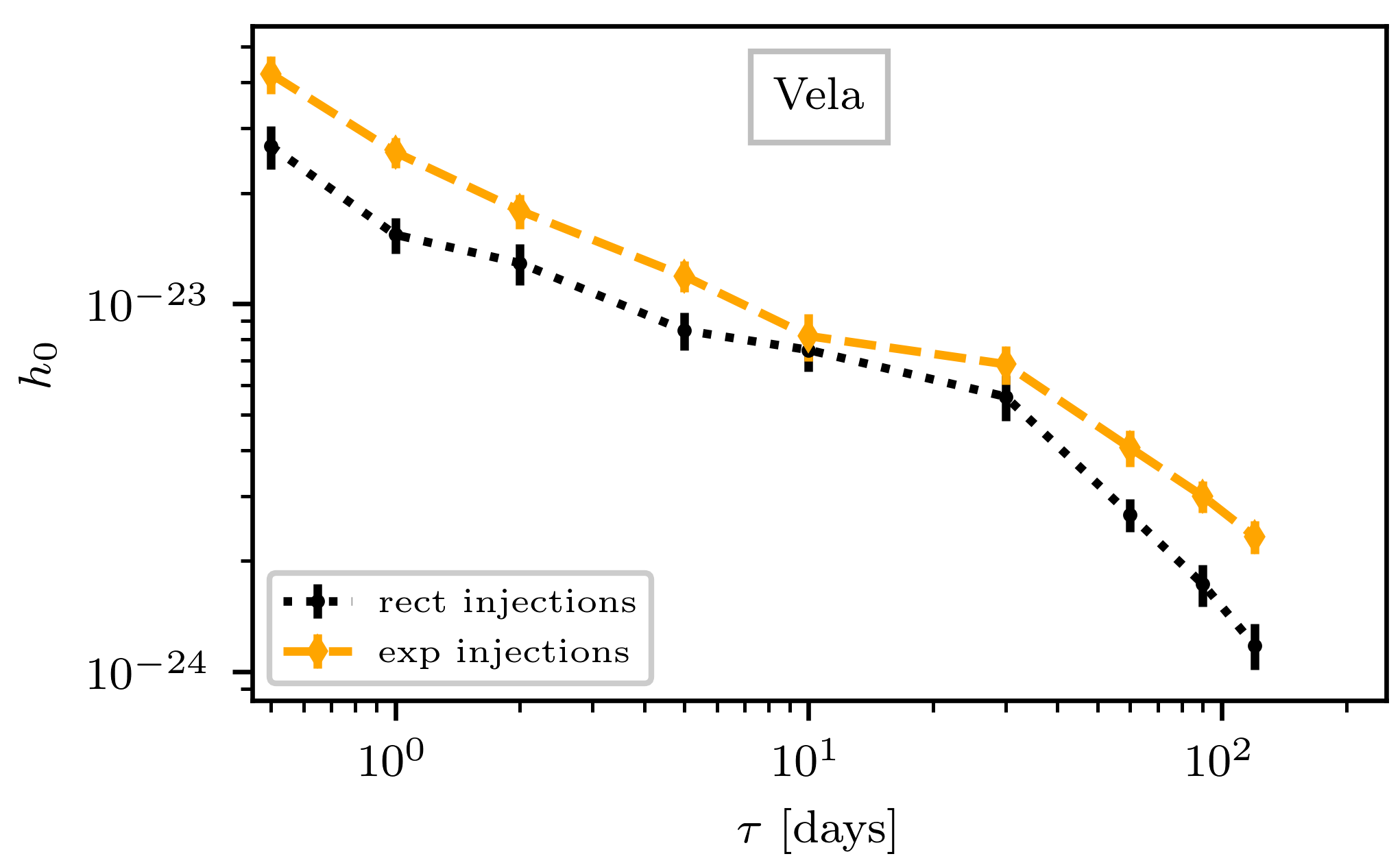}
 \caption{\label{fig:vela-exp-rect}
  Injection recovery results for the Vela search,
  comparing transient injections with
  rectangular window function
  (black dotted line,
   same as in Fig.~\ref{fig:uls-tau})
  and exponential window function
  (orange dashed line),
  both recovered with rectangular search windows.
 }
\end{figure}

This question was previously addressed in Sec. V.B of~\cite{Prix:2011qv}
through looking at
detection probability
$\pdet$
against
false-alarm probability
$\pfa$
for synthetic noise and signal draws.
They found moderate losses
with a worst case of $\lesssim10\%$ at relatively high $\pfa$.

As a specific test with more direct relation to the analysis in this paper,
let us revisit the \ul injection-and-recovery procedure
with exponential-window injections.
Direct comparison of recovering these injections
with both rectangular and exponential search windows
is too expensive for the present purpose;
instead, we have simply tested
the recovery of exponential injections with rectangular search windows --
though only for the 20161212 Vela glitch as an example target.
Results are shown in Fig.~\ref{fig:vela-exp-rect},
showing that there is indeed some loss in sensitivity
from transient window mismatch,
and that a dedicated exponential-window search
(likely on \gpu[s]~\cite{Keitel:2018pxz})
can be valuable in the future.
But for the moment,
this test also demonstrates that our simplified pilot search
also had sensitivity to exponentially decaying signals,
with slightly higher \ul[s] as per Fig.~\ref{fig:vela-exp-rect}.

It is worth noting that,
following the conventions of~\cite{Prix:2011qv},
the length of an exponential signal with
\mbox{$h_0(t)\propto\exp[-(t - t_0)/\tau]$}
is longer than
that of a rectangular window with fixed length $\tau$:
the initial amplitude falls to
37\% after $1\tau$,
14\% after $2\tau$
and 5\% after $3\tau$,
at which point the LALSuite implementation cuts off the window.
Hence, one cannot directly compare $\pdet$ at a fixed $\tau$
for injections of both window types and purely attribute the difference to window mismatch.

\bibliography{biblio}

\end{document}